\documentclass[11pt,twoside]{article}
\usepackage{qic}
\usepackage{bbm}
\usepackage{color}
\usepackage{amsmath}
\usepackage{latexsym}
\newtheorem{thm}{Theorem}
\newtheorem{remark}{Remark}

\newcommand{\proofmod}[2]{{\bf #1.} #2 $\Box$.}

\begin{document}
\setlength{\textheight}{8.0truein}    

\runninghead{Limitation for linear maps $\ldots$}
            {A. SaiToh, R. Rahimi, and M. Nakahara}

\normalsize\textlineskip
\thispagestyle{empty}
\setcounter{page}{1}

\copyrightheading{0}{0}{2003}{000--000}

\vspace*{0.88truein}

\alphfootnote

\fpage{1}

\centerline{\bf
Limitation for linear maps in a class for detection and}
\vspace*{0.035truein}
\centerline{\bf quantification of bipartite nonclassical correlation}
\vspace*{0.37truein}
\centerline{\footnotesize
AKIRA SAITOH$^{1}$\fnm{$*$}\fnt{$*$}{Present address:
Quantum Information Science Theory Group, National Institute of Informatics,
2-1-2 Hitotsubashi, Chiyoda, Tokyo 101-8430, Japan}\fnt{$*$}{E-mail address: akira@m.ieice.org},
~ROBABEH RAHIMI$^{2}$\fnm{$\dagger$}\fnt{$\dagger$}{E-mail address: rrahimid@uwaterloo.ca},
~MIKIO NAKAHARA$^{1,3}$\fnm{$\ddagger$}\fnt{$\ddagger$}{E-mail address: nakahara@math.kindai.ac.jp}}
\vspace*{0.015truein}
\centerline{\footnotesize\it
$^1$Research Center for Quantum Computing, Interdisciplinary}
\baselineskip=10pt
\centerline{\footnotesize\it
Graduate School of Science and Engineering, Kinki University}
\baselineskip=10pt
\centerline{\footnotesize\it 3-4-1 Kowakae, Higashi-Osaka, Osaka 577-8502, Japan}
\centerline{\footnotesize\it
$^2$Institute for Quantum Computing, University of Waterloo}
\baselineskip=10pt
\centerline{\footnotesize\it
200 University Avenue West, Waterloo, ON N2L 3G1, Canada}
\centerline{\footnotesize\it
$^3$Department of Physics, Kinki University}
\baselineskip=10pt
\centerline{\footnotesize\it 3-4-1 Kowakae, Higashi-Osaka, Osaka 577-8502, Japan}
\vspace*{0.225truein}
\publisher{(received date)}{(revised date)}
\vspace*{0.21truein}

\abstracts{
Eigenvalue-preserving-but-not-completely-eigenvalue-preserving (EnCE)
maps were previously introduced for the purpose of detection
and quantification of nonclassical correlation, employing the
paradigm where nonvanishing quantum discord implies the existence of
nonclassical correlation. It is known that only the matrix transposition
is nontrivial among Hermiticity-preserving (HP) linear EnCE maps when
we use the changes in the eigenvalues of a density matrix due to a partial map
for the purpose. In this paper, we prove that this is true even among
not-necessarily HP (nnHP) linear EnCE maps. The proof utilizes a
conventional theorem on linear preservers. This result imposes a strong
limitation on the linear maps and promotes the importance of nonlinear
maps.
}{}{}

\vspace*{10pt}

\vspace*{3pt}
\communicate{to be filled by the Editorial}

\vspace*{1pt}\textlineskip    

\section{Introduction}
Nonclassical correlation in a bipartite quantum system defined
differently from entanglement has been studied in several different
contexts \cite{B99,Z02,O02} and attracting much interest in quantum
information science. There are quantum information processing schemes,
such as superdense coding \cite{BW92,R06} and
some of quantum games \cite{jS04,OjS04,SRNgame09,NT10},
having advantages over classical counterparts even with a slightly
polarized pseudo-entangled state that possesses a certain amount of
nonclassical correlation.
It is also known that the quantum algorithm for an approximate trace
estimation of a unitary matrix using a single pseudo-pure qubit and 
other qubits in the maximally mixed state \cite{KL98} involves no
entanglement but a considerable amount of nonclassical correlation
\cite{DSC08}. With these examples, it is highly motivated to study
well-formulated nonclassical correlation.

A bipartite density matrix $\rho^{\rm AB}$ is (properly) classically
correlated (pcc) \cite{O02,H05} if and only if it has a product eigenbasis:
\[
 \rho_{\rm pcc}^{\rm AB}=\sum_{ij=11}^{d^{\rm A}d^{\rm B}}
e_{ij}|u_i\rangle^{\rm A}\langle u_i|
\otimes |v_j\rangle^{\rm B}\langle v_j|
\]
where $e_{ij}$ are the eigenvalues; $|u_i\rangle$ and $|v_j\rangle$ are
the eigenvectors of the reduced density matrices of subsystems
${\rm A}$ and ${\rm B}$, respectively;
$d^{\rm A}$ and $d^{\rm B}$ are the dimensions of the Hilbert spaces of
the respective subsystems. Any $\rho^{\rm AB}$ having no product eigenbasis is
nonclassically correlated (ncc). These definitions are equivalent to
regarding a bipartite quantum state with a nonvanishing quantum discord
(for any side) \cite{Z02} as a nonclassically correlated state and a
state with a vanishing quantum discord as a classically correlated
state. Here, we have used the term quantum discord as the one defined with
a minimization.

There have been several detection methods and measures of nonclassical
correlation for this paradigm; some of them are computationally expensive
while perfect in the detection range \cite{Z02,O02,PHH08,SRN08,LF10}
(there are some conditionally\footnote{
To be tractable, the measures in Refs.\ \cite{G07,L08}
require that a proper Schmidt basis be found within polynomial time in
$d^{\rm A}d^{\rm B}$.}~
computationally tractable ones \cite{G07,L08});
the others are computationally tractable while imperfect in the
detection range \cite{SRN11,SRN10-2}. Recently, Daki\'{c} {\em et
al.} \cite{DVB10} proposed a computationally tractable method detecting
nonclassical correlation perfectly using an operator Schmidt decomposition.
Chen {\em et al.} \cite{CCMV11} also proposed a computationally tractable
method\footnote{They proved that
a necessary and sufficient condition for a bipartite density matrix
$\rho^{\rm AB}$ to be a one-way classically correlated (1wcc) state
$\rho^{\rm AB}_{\rm 1wcc}
=\sum_j\sigma^{\rm A}_j\otimes |v_j\rangle^{\rm B}\langle v_j|$
($\sigma^{\rm A}_j$'s are positive semidefinite Hermitian matrices
acting of subsystem ${\rm A}$ and $\{|v_j\rangle^{\rm B}\}$ is a complete
orthonormal set (CONS) of subsystem ${\rm B}$) is that,
for an arbitrary CONS $\{|u_i\rangle^{\rm A}\}$ ($i=1,\ldots,d^{\rm
A}$) of subsystem ${\rm A}$, all ${}^{\rm A}\langle u_i|\rho^{\rm
AB}|u_{i'}\rangle^{\rm A}$'s commute with each other ($i,i'=1,\ldots,d^{\rm A}$).
By their theorem, one has only to test a polynomial number of
commuting relations (by applying their theorem to both sides) to
decide whether $\rho^{\rm AB}$ has a product eigenbasis or not.}~
to detect nonclassical correlation perfectly.

Among tractable albeit imperfect ones, the scheme we proposed
in Ref.~\cite{SRN11} uses a certain class of maps acting on a given
density matrix. This scheme is in analogy with the scheme using
positive-but-not-completely-positive (PnCP) maps \cite{P96,H96,H97}
commonly used for entanglement detection and quantification.
It has been of our interest to find certain maps that are useful for
detecting and quantifying nonclassical correlation of a quantum
bipartite system. There have been some classes of maps already defined
in our previous works \cite{SRN08-2,SRN11} for this purpose.

Let us first introduce those classes of maps.
A map that preserves the eigenvalues and their algebraic multiplicities
of a matrix is called an eigenvalue-preserving (EP) map.
Sometimes the term ``spectrum-preserving'' is used in the same meaning,
but it often means spectrum-preserving without taking the 
multiplicity into account. Therefore, we employ the term EP in this paper.
Among EP maps, Hermiticity preserving (HP) maps were thought to be rather
natural in quantum physics. We considered linear and nonlinear HP EP
maps that are not completely EP in our previous contribution
\cite{SRN11}. We call such maps HP
eigenvalue-preserving-but-not-completely-eigenvalue-preserving (EnCE)
maps.

An HP EP map $\Lambda_{\rm HPEP}$ maps a density matrix $\rho$ to a
density matrix $\rho'$ with the same eigenvalues by definition. Thus
\[
 \Lambda_{\rm HPEP}: \sum_{i}e_i|v_i\rangle\langle v_i|
\mapsto \sum_{i}e_i|{v_i}'\rangle\langle {v_i}'|
\]
with $e_i$ the $i$th eigenvalue of $\rho$ and $\rho'$;
$|v_i\rangle$ and $|{v_i}'\rangle$ the corresponding eigenvectors of
$\rho$ and $\rho'$, respectively. Therefore, $\Lambda_{\rm HPEP}$ maps any
projector set $\{|v_i\rangle\langle v_i|\}$ corresponding to a complete
orthonormal set (CONS) $\{|v_i\rangle\}$ to another projector set
$\{|{v_i}'\rangle\langle {v_i}'|\}$ corresponding to a CONS $\{|{v_i}'\rangle\}$.

An HP EnCE map is an HP EP map that is not completely EP,
i.e., $I^{\rm A}\otimes \Lambda_{\rm HPEP}^{\rm B}$ does not always
preserve the eigenvalues of a density matrix for some dimensions of
subsystem ${\rm A}$. For any linear or nonlinear HP EnCE map
$\Lambda_{\rm HPEnCE}$, the maps $I^{\rm A}\otimes \Lambda_{\rm HPEnCE}^{\rm B}$
and $\Lambda_{\rm HPEnCE}^{\rm A}\otimes I^{\rm B}$ both
preserve the eigenvalues of a density matrix $\rho^{\rm AB}$ when it
has a product eigenbasis, by the definition of the class of the maps,
while they may change when it has no product eigenbasis.
Therefore, changes in the eigenvalues under the action of
$I^{\rm A}\otimes \Lambda_{\rm HPEnCE}^{\rm B}$ and/or
$\Lambda_{\rm HPEnCE}^{\rm A}\otimes I^{\rm B}$ can be used for
detection and quantification of nonclassical correlation.
As an example of detection, consider the state $\rho_p = (1-p)I/d+p\rho_{\rm NPT}$
with $0< p\le 1$, $d$ the dimension of the Hilbert space, and $\rho_{\rm NPT}$ a
state possessing negative partial transposition (NPT) \cite{P96,H96}, and consider
the matrix transposition $\mathcal{T}$ as an HP EnCE map. (The state $\rho_p$ is
possibly a separable state.) Obviously, $(I\otimes\mathcal{T})\rho_p$ has different
eigenvalues from $\rho_p$. As for quantification, 
we introduced a measure using a form of logarithmic fidelity,
which is subadditive when an HP EnCE map is chosen appropriately \cite{SRN11}.

In particular for HP linear EnCE maps, the only nontrivial map is
$\mathcal{T}$ as we showed in Ref.~\cite{SRN11} using Wigner's
unitary-antiunitary theorem \cite{W59,Wig,Wig2,SA90}. One may notice
that an HP linear EnCE map is nothing but an operation to change an
orthonormal basis to another one with the same dimension. In short, any
HP linear EP map can be decomposed into unitary transformations and
$\mathcal{T}$. Therefore, any HP linear EnCE map can also be decomposed
into unitary transformations and $\mathcal{T}$. To overcome this
limitation, we introduced a nonlinear extension in the previous work \cite{SRN11}.

A natural question arises as to whether there is a nontrivial linear
EnCE map other than $\mathcal{T}$ if we consider not-necessarily
Hermiticity preserving (nnHP) maps. This contribution will reach
a negative answer with the following theorem.
\begin{thm}\label{thmmain}
For any nnHP linear EnCE map $\Upsilon$ and an identity map $I$ of any
dimension, the set of eigenvalues of $(I^{\rm A}\otimes \Upsilon^{\rm B})\rho^{\rm AB}$
is the same as that of $\rho^{\rm AB}$ for all $\rho^{\rm AB}$, or the same
as that of $(I^{\rm A}\otimes \mathcal{T}^{\rm B})\rho^{\rm AB}$ for all
$\rho^{\rm AB}$, where $\rho^{\rm AB}$ is a density matrix of a bipartite system
${\rm AB}$.
\end{thm}

This is the main theorem we are going to prove.
It suggests that, even among nnHP linear EnCE maps,
$\mathcal{T}$ is the only nontrivial map for the purpose of detecting
and quantifying nonclassical correlation as far as we use the changes in
the eigenvalues. Thus, linear ones cannot detect and quantify
nonclassical correlation of the states whose eigenvalues are unchanged
by the partial transposition, such as one-way classically correlated
states \cite{H05}.

The result shown here should not be regarded as a
limitation for general linear operations although it is, of course,
a natural choice to explore nonlinear EnCE maps.

In this paper, we begin with preliminary lemmas in Section~\ref{secPre} and
describe the results in the subsequent sections.
As we mentioned, we are going to prove that $\mathcal{T}$ is the only
nontrivial nnHP linear EnCE map for our purpose. This will be proved
in two different contexts.
In Section\ \ref{secMinc}, we discuss the case where we regard a map as
one preserving eigenvalues of any matrix. In Section\ \ref{secMain},
we discuss the case where we regard a map as one preserving
eigenvalues of any density matrix. The results are summarized in
Section\ \ref{secConc} with remarks.

\section{Preliminary Lemmas}\label{secPre}
Let us begin with two lemmas for linear maps that
are used in proofs hereafter.
Here, $M_d(\mathbbm{F})$ denotes the set of $d\times d$ matrices over
a field $\mathbbm{F}$ and $\mathbbm{S}_d$ denotes the set of
$d\times d$ density matrices (positive semidefinite Hermitian matrices
with unit trace).
In addition, as usual, linearity is defined in the following way.
\begin{definition}\label{deflin}
For two spaces $V$ and $W$ of matrices over $\mathbbm{F}$, we say
that a map $\Lambda: V\rightarrow W$ is linear if
$\Lambda(\alpha u + \beta v)=\alpha\Lambda(u)+\beta\Lambda(v)$
for all $\alpha,\beta\in\mathbbm{F}$ and $u,v\in V$.
\end{definition}

Now the lemmas are stated as follows.
\begin{lemma}\label{lemma1}
Consider linear maps $\Lambda_1: M_d(\mathbbm{C})\rightarrow
M_d(\mathbbm{C})$ and $\Lambda_2: M_d(\mathbbm{C})
\rightarrow M_d(\mathbbm{C})$ with integer $d>0$.
Suppose they are equivalent as maps, i.e., $\Lambda_1 L = \Lambda_2 L$
for all $L\in M_d(\mathbbm{C})$. Then, $I\otimes\Lambda_1$
and $I\otimes\Lambda_2$ are equivalent as maps for an identity map $I$
acting on the space of $q\times q$ matrices for any integer $q>1$, i.e.,
$(I\otimes\Lambda_1)M=(I\otimes\Lambda_2)M$ for all
$M\in M_{qd}(\mathbbm{C})$.
\end{lemma}
\proof{
By linearity, for any matrix $M=\sum_{i,j,k,l=1,1,1,1}^{q,q,d,d}
c_{ijkl}|i\rangle\langle j|\otimes |k\rangle\langle l|$ with
$c_{ijkl}\in\mathbbm{C}$, we have
$(I\otimes\Lambda_1)M=\sum_{ijkl}c_{ijkl}|i\rangle\langle j|\otimes
\Lambda_1(|k\rangle\langle l|)$ and
$(I\otimes\Lambda_2)M=\sum_{ijkl}c_{ijkl}|i\rangle\langle j|\otimes
\Lambda_2(|k\rangle\langle l|)$.
Obviously, $(I\otimes\Lambda_1)M=(I\otimes\Lambda_2)M$ since, by the
assumption, $\Lambda_1(|k\rangle\langle l|)=\Lambda_2(|k\rangle\langle
 l|)$ $\forall k,l$.
}
\begin{lemma}\label{lemma2}
Consider linear maps $\Lambda_1: \mathbbm{S}_d\rightarrow
M_d(\mathbbm{C})$ and $\Lambda_2: \mathbbm{S}_d
\rightarrow M_d(\mathbbm{C})$ with integer $d>0$.
Suppose $\Lambda_1 \rho = \Lambda_2 \rho$ for all
$\rho\in\mathbbm{S}_d$.  Then,
$(I\otimes\Lambda_1)\sigma=(I\otimes\Lambda_2)\sigma$
for an identity map $I$ acting on the space of $q\times q$ density matrices 
for any integer $q>1$ and all $\sigma\in\mathbbm{S}_{qd}$.
\end{lemma}
\proof{
It is always possible to represent a $qd\times qd$ density matrix
$\sigma$ as $\sigma=\sum_{ij}c_{ij}\alpha_i\otimes\beta_j$ with
a finite number of
$c_{ij}\in\mathbbm{R}$ (note that they may be negative),
$q\times q$ density matrices $\alpha_i$, and $d\times d$ density
matrices $\beta_j$. This is because any Hermitian basis operator can
be represented as a linear combination of a finite number of density
matrices. By linearity of $\Lambda_1$ and $\Lambda_2$,
$(I\otimes\Lambda_1)\sigma=\sum_{ij}c_{ij}\alpha_i\otimes\Lambda_1\beta_j$
and
$(I\otimes\Lambda_2)\sigma=\sum_{ij}c_{ij}\alpha_i\otimes\Lambda_2\beta_j$
hold. By the assumption, $\Lambda_1\beta_j=\Lambda_2\beta_j$ $\forall j$. Hence
$(I\otimes\Lambda_1)\sigma=(I\otimes\Lambda_2)\sigma$.
}

\section{A Theorem on Linear Preservers Revisited}\label{secMinc}
The maps we consider are those preserving the eigenvalues
of a density matrix. In this regard, it is essential to revisit the
long-standing research on linear preservers \cite{Mar71,LT92,LP01}
in the theory of linear algebra. When we consider the linear maps preserving
eigenvalues of any $d\times d$ matrix, then it is rather immediate
to show that the transposition is the only nontrivial map in the
context, as shown in this section. However, it should be
noted that the maps of our interest are those preserving the eigenvalues
of the density matrices; thus the class of the maps are possibly
different from those acting on the space of general square matrices. Some
additional mathematical evaluations are required as we will describe in
the next section for the maps of our interest.

Let us revisit the result of the Marcus-Moyls theorem \cite{MM59}
and Minc's theorem \cite{Minc74,Minc88} on linear transformations.
\begin{thm}[From Marcus and Moyls, 1959 and Minc, 1974]\label{ThMMM}
A linear transformation $\Lambda$ on $M_d(\mathbbm{C})$
preserves the determinant and the trace for all
$A\in M_d(\mathbbm{C})$ if and only if $\Lambda$ preserves the
eigenvalues (including their algebraic multiplicities) for all
$A\in M_d(\mathbbm{C})$. Then there exists a matrix $S$ in
$M_d(\mathbbm{C})$ such that
$\Lambda A=S^{-1}AS$ for all $A\in M_d(\mathbbm{C})$, or
$\Lambda A=S^{-1}A^TS$ for all $A\in M_d(\mathbbm{C})$.
\end{thm}
\proof{
This is a direct consequence of the Marcus-Moyls theorem \cite{MM59}
and Minc's theorem \cite{Minc74,Minc88} on linear preservers.
}~

Suppose we require a linear map $\Lambda$ acting on the space of
$d^{\rm B}\times d^{\rm B}$ matrices to preserve the eigenvalues of
any matrix in $M_{d^{\rm B}}(\mathbbm{C})$ rather than density matrices only.
Then, by Theorem \ref{ThMMM} and Lemma \ref{lemma1},
$I^{\rm A}\otimes\Lambda^{\rm B}$ can be written as
\[
 I^{\rm A}\otimes\Lambda^{\rm B}: A\mapsto
[I^{\rm A}\otimes (S^{\rm B})^{-1}]A'(I^{\rm A}\otimes S^{\rm B})
\]
where $A'$ stands for either $A$ for all $A\in M_{d^{\rm A}d^{\rm B}}(\mathbbm{C})$
or $(I^{\rm A}\otimes \mathcal{T}^{\rm B})A$
(namely, the partial transposition of $A$)
for all $A\in M_{d^{\rm A}d^{\rm B}}(\mathbbm{C})$.
Consequently, the set of eigenvalues
of $(I^{\rm A}\otimes\Lambda^{\rm B})A$ is equal to either
that of $A$ or that of $(I^{\rm A}\otimes \mathcal{T}^{\rm B})A$.
The partial transposition
is thus the only nontrivial linear map as far as we use the changes in
the eigenvalues for our purpose.

As we have already mentioned, this result is not directly applicable to the
case where $\Lambda$ is required to preserve the eigenvalues for
density matrices only and not necessarily for general matrices.
(In addition, we do not assume that Hermiticity is preserved in general.)

\section{Formal Definitions and the Proof for the Main Theorem}\label{secMain}
In this section, we are going to prove our main theorem (Theorem~\ref{thmmain})
step by step. Due to its context, we restrict ourselves within maps from the set
of density matrices to the set of square matrices. Note that the image need
not be Hermitian.

We begin with the definitions of maps of our interest without imposing
linearity.
\begin{definition}\label{defnnHPEP}
A not-necessarily Hermiticity preserving (nnHP) eigenvalue-preserving
(EP) map $\Theta$ is a map from $\mathbbm{S}_d$ to $M_d(\mathbbm{C})$ such that, for a
$d\times d$ density matrix $\rho = \sum_{i=1}^d e_i|v_i\rangle\langle v_i|$ ($e_i$
and $|v_i\rangle$ are the $i$th eigenvalue and the corresponding eigenvector),
$\Theta: \rho\mapsto A\in M_d(\mathbbm{C})$
with a matrix $A$ having the set of eigenvalues same as that of
$\rho$, for all $\rho\in\mathbbm{S}_d$.
\end{definition}
\begin{definition}
An nnHP EP-but-not-Completely-EP (nnHP EnCE) map is an nnHP EP map
$\tilde\Theta$ such that $I\otimes \tilde\Theta$ for some dimension
for $I$ does not preserve eigenvalues of some density matrix.
\end{definition}

When we impose linearity (see Definition \ref{deflin}), the class of
nnHP EP maps has an interesting property as stated in the following
theorem. The proof is given later in this section.
\begin{thm}\label{varMinc}
For an nnHP linear EP map
$\Theta_{\rm lin}:\mathbbm{S}_d \rightarrow M_d(\mathbbm{C})$,
there exists a matrix $S$ in
$M_d(\mathbbm{C})$ such that either
$\Theta_{\rm lin} \rho=S^{-1}\rho S$ for all $\rho\in \mathbbm{S}_d$ or
$\Theta_{\rm lin} \rho=S^{-1}\rho^T S$ for all $\rho\in \mathbbm{S}_d$.
\end{thm}

(Note: In case we also impose the HP property, then $S$ becomes unitary;
the theorem under this restriction is equivalent to the statement of
Proposition 2 of our previous work \cite{SRN11}.)
The theorem suggests that $\Theta_{\rm lin}$ must map
the set $\{|v_i\rangle\langle v_i|\}$ of any $\rho$ to some set
$\{|a_i\rangle\langle b_i|\}$ where vector sets
$\{|a_i\rangle\}$ and $\{|b_i\rangle\}$ are biorthogonal\footnote{
Generally, we say that two systems of elements
$\{x_1,\ldots,x_d\}$ and $\{y_1,\ldots,y_d\}$ in a unitary space are
biorthogonal if $(x_i, y_j)=\delta_{ij}$ $(1\le i,j \le d)$ where
$(~,~)$ is an inner product (see, e.g., Ref.\ \cite{LT85}).
In the present context, of course, a standard inner product is
employed.}~, i.e., the left and right eigenvectors of $\Theta_{\rm lin}\rho$
are $(|b_i\rangle)^*$ and $|a_i\rangle$, respectively.

Before going into the proof of the theorem, let us briefly
overview how such a map is related to detection and quantification of
nonclassical correlation. For any nnHP linear EnCE map 
$\Upsilon$, we have
\[
 (I^{\rm A}\otimes {\Upsilon}^{\rm B})\rho_{\rm pcc}^{\rm AB}
=\sum_{ij}e_{ij}|u_i\rangle^{\rm A}\langle u_i|\otimes
|a_j\rangle^{\rm B}\langle b_j|.
\]
The eigenvalues of $\rho_{\rm pcc}^{\rm AB}$ are preserved by
$I^{\rm A}\otimes {\Upsilon}^{\rm B}$ (and by
${\Upsilon}^{\rm A}\otimes I^{\rm B}$).
Thus changes in the eigenvalues of $\rho^{\rm AB}$ by
$I^{\rm A}\otimes {\Upsilon}^{\rm B}$ or
by ${\Upsilon}^{\rm A}\otimes I^{\rm B}$
imply that $\rho^{\rm AB}$ has no product eigenbasis.
The original and the (possibly) transformed eigenvalues are used to
define a measure of nonclassical correlation as we discussed in
Ref.~\cite{SRN11}.

To prove Theorem \ref{varMinc}, we firstly introduce a remark, several
lemmas, and the Marcus-Moyls theorem \cite{MM59}. Here, $\mathbbm{H}_d$
denotes the set of $d\times d$ Hermitian matrices.
\begin{remark}
An nnHP linear EP map $\Theta_{\rm lin}$ preserves the set of
eigenvalues of any positive semidefinite Hermitian matrix $A_+$.
\end{remark}

This is obvious considering the fact that $A_+$ is proportional to
some density matrix or otherwise a zero matrix, and $\Theta_{\rm lin}$
is linear. The proofs hereafter are tacitly based on this remark.
We now introduce the lemmas.
\begin{lemma}\label{lemmaL0}
An nnHP linear EP map $\Theta_{\rm lin}$ can be regarded as a map
from $\mathbbm{H}_d$ to $M_d(\mathbbm{C})$
such that, for all $A\in \mathbbm{H}_d$, the set of eigenvalues of
$\Theta_{\rm lin}(A)$ is equal to that of $A$, if 
$\Theta_{\rm lin}$ is unital (i.e., if $\Theta_{\rm lin}(I)=I$).
\end{lemma}
\proof{
By linearity, $\Theta_{\rm lin}(A)=\Theta_{\rm lin}[(A+cI)-(cI)]
=\Theta_{\rm lin}(A+cI)-c\Theta_{\rm lin}(I)$
for some $c\in\mathbbm{R}$ such that $A+cI\ge 0$.
By assumption, $\Theta_{\rm lin}$ is unital. Therefore
$\Theta_{\rm lin}(A)=\Theta_{\rm lin}(A+cI)-cI$. By the fact that
the eigenvalues of $\Theta_{\rm lin}(A+cI)$ are same as
those of $A+cI$ by definition, the proof is now completed.
}~

We can prove that any nnHP linear EP map is in fact a unital map.
\begin{lemma}\label{lemmaL1}
Any nnHP linear EP map $\Theta_{\rm lin}$ satisfies $\Theta_{\rm lin}(I)=I$.
\end{lemma}
\proof{
Consider a density matrix $\rho\in\mathbbm{S}_d$ with eigenvalues
$e_1,\ldots, e_d$. By definition, $\Theta_{\rm lin}$ preserves the
eigenvalues of $\rho+cI$ for any $c\ge -{\rm min}_k e_k$ ($k=1,\ldots,d$).
Consider such $c$. We have ${\rm det}\,\Theta_{\rm lin}(\rho+cI) = \prod_k(e_k+c)$.
Now consider the matrix $\Theta_{\rm lin}(I)^{-1}$, namely, the inverse matrix
of $\Theta_{\rm lin}(I)$. This has the eigenvalue
$1$ with multiplicity $d$. Thus ${\rm det}\,\Theta_{\rm lin}(I)^{-1}=1$.
We have ${\rm det}\,\Theta_{\rm lin}(\rho+cI)=
{\rm det}\,[\Theta_{\rm lin}(\rho+cI)\Theta_{\rm lin}(I)^{-1}]
={\rm det}\,[\Theta_{\rm lin}(\rho)\Theta_{\rm lin}(I)^{-1}+cI]$
by linearity of $\Theta_{\rm lin}$. Let us write the eigenvalues of
$\Theta_{\rm lin}(\rho)\Theta_{\rm lin}(I)^{-1}$ as
$\lambda_1,\ldots,\lambda_d$. Then we have
$\prod_{k}(\lambda_k+c)=\prod_k(e_k+c)$~~$\forall c\ge -{\rm min}_k e_k$.
Therefore, $\lambda_k=e_k$ for $k=1,\ldots,d$. Thus, the matrix
$\Theta_{\rm lin}(\rho)\Theta_{\rm lin}(I)^{-1}$ has the same
eigenvalues as those of $\rho$.
Consider the case where $\rho$ has mutually distinct nonzero eigenvalues.
Then, $\Theta_{\rm lin}(\rho)\Theta_{\rm lin}(I)^{-1}$ is a nonsingular simple
matrix; hence there exists nonsingular $Q\in M_d(\mathbbm{C})$ such that
$\Theta_{\rm lin}(\rho)\Theta_{\rm lin}(I)^{-1}=QDQ^{-1}$ with nonsingular
$D={\rm diag}(e_1,\ldots,e_d)$. We then use the fact that
$\Theta_{\rm lin}(I)\Theta_{\rm lin}(\rho)\Theta_{\rm lin}(I)^{-1}$
is a similarity transformation of $\Theta_{\rm lin}(\rho)$, and
hence it is a similarity transformation of $D$.  It follows that
$\Theta_{\rm lin}(I)QDQ^{-1}$ is a similarity transformation of $D$. Now
we use the fact that the Jordan decomposition of $\Theta_{\rm lin}(I)$ is
given by $\Theta_{\rm lin}(I) = I + N$ with $N$ the nilpotent term\footnote{
Because the eigenvalue of $\Theta_{\rm lin}(I)$ is $1$ with multiplicity $d$,
we have the Jordan decomposition $\Theta_{\rm lin}(I) = W I W^{-1} + N
= I + N$ for some nonsingular $W\in M_d(\mathbbm{C})$.}.
We have $\Theta_{\rm lin}(I)QDQ^{-1} = QDQ^{-1} + NQDQ^{-1}$.
Because this is similar to $D$, its Jordan decomposition is written
as $\tilde S + \tilde N$ with $\tilde S$ similar to $D$ and
$\tilde N=0$. Hence the term $NQDQ^{-1}$ should vanish, which
implies $N=0$. Therefore, $\Theta_{\rm lin}(I)=I$ holds.
}~

With Lemmas \ref{lemmaL0} and \ref{lemmaL1}, we can state the following
lemma.  
\begin{lemma}\label{lemma3-1}
An nnHP linear EP map $\Theta_{\rm lin}$ can be regarded as a map
from $\mathbbm{H}_d$ to $M_d(\mathbbm{C})$
such that, for all $A\in \mathbbm{H}_d$, the set of eigenvalues of
$\Theta_{\rm lin}(A)$ is equal to that of $A$.
\end{lemma}
\proof{
This is a consequence of Lemmas \ref{lemmaL0} and \ref{lemmaL1}.
}~

We introduce one more lemma.
\begin{lemma}\label{lemma3-2}
A linear map $\Pi:\mathbbm{H}_d \rightarrow M_d(\mathbbm{C})$ can be
regarded as a linear map on
$M_d(\mathbbm{C})$ [namely, a linear map from $M_d(\mathbbm{C})$
to $M_d(\mathbbm{C})$].
\end{lemma}
\proof{
For all $A\in M_d(\mathbbm{C})$, there exists a Cartesian decomposition
$A=H_1+iH_2$ with $i=\sqrt{-1}$, $H_1=(A+A^\dagger)/2$, and
$H_2=(A-A^\dagger)/(2i)$ (See, e.g., pp.178-179 of Ref.\ \cite{LT85}).
Because $H_1$ and $H_2$ are Hermitian, we have $\Pi(A)=\Pi(H_1)+i\Pi(H_2)$. 
}~

In addition to the above lemmas, it is essential to introduce a useful
theorem by Marcus and Moyls \cite{MM59}.
\begin{thm}[Marcus and Moyls, 1959]\label{theorem3-3}
Let $\Lambda$ be a linear map on $M_d(\mathbbm{C})$. The following
conditions are equivalent:\\
(i) $\Lambda$ preserves eigenvalues for all Hermitian matrices
in $M_d(\mathbbm{C})$.\\
(ii) $\Lambda$ preserves eigenvalues for all matrices in
 $M_d(\mathbbm{C})$.\\
(iii) There exists a unimodular matrix $L$ such that either
$\Lambda A=LAL^{-1}$ for all $A\in M_d(\mathbbm{C})$ or 
$\Lambda A=LA^TL^{-1}$ for all $A\in M_d(\mathbbm{C})$.
\end{thm}
\proof{
The proof is found in Ref.\ \cite{MM59}.
}~

With the above preparation, we can now prove Theorem \ref{varMinc}.\\
\proofmod{Proof of Theorem \ref{varMinc}}{
By Lemmas\ \ref{lemma3-1} and \ref{lemma3-2}, an nnHP linear EP map
$\Theta_{\rm lin}$ can be regarded as a linear map on $M_d(\mathbbm{C})$
that preserves the eigenvalues of any Hermitian matrix.
By Theorem\ \ref{theorem3-3}, the proof is completed.
}~

With Theorem\ \ref{varMinc} and Lemma\ \ref{lemma2}, we
obtain our main theorem:
\setcounter{thm}{0}
\begin{thm}
For any nnHP linear EnCE map $\Upsilon$ and a bipartite system
${\rm AB}$, the set of eigenvalues of $(I^{\rm A}\otimes
 \Upsilon^{\rm B})\rho^{\rm AB}$ is equal to that of
$\rho^{\rm AB}$ for all $\rho^{\rm AB}\in\mathbbm{S}_{d^{\rm A}d^{\rm B}}$
or equal to that of $(I^{\rm A}\otimes \mathcal{T}^{\rm B})\rho^{\rm
 AB}$ for all $\rho^{\rm AB}\in\mathbbm{S}_{d^{\rm A}d^{\rm B}}$,
where $\mathcal{T}$ is the matrix transposition.
\end{thm}
\proof{Theorem\ \ref{varMinc} and Lemma\ \ref{lemma2} prove the theorem.}~

Besides our main theorem, Definition\ \ref{defnnHPEP}, Lemma\ \ref{lemma3-1}, and
Lemma\ \ref{lemma3-2} also imply the following fact.
\begin{remark}
The following condition is equivalent to conditions (i)-(iii) of
Theorem\ \ref{theorem3-3}.\\
(iv) $\Lambda$ preserves eigenvalues for all density matrices in $M_d(\mathbbm{C})$.
\end{remark}

\section{Concluding Remarks}\label{secConc}
We have proved that, for a linear map $\Upsilon$ in the class of nnHP
EnCE and a bipartite density matrix $\rho$, $(I\otimes\Upsilon)\rho$ has
the same set of eigenvalues as $\rho$ or its partial transposition.
This indicates that nontrivial nnHP EnCE maps for detection and
quantification of nonclassical correlation, except for the
transposition, should be nonlinear.

To achieve this result, we have used a conventional theorem in the
theory of linear preservers (See, e.g., Refs.\ \cite{Mar71,LT92,LP01}).
Linear maps locally acting on a quantum bipartite system are often found
useful to characterize nonclassicality of correlation, such as PnCP maps
used in the entanglement theory. In this sense, it is expected that more
conventional theorems will be found to be useful in the development of
the theory of bipartite nonclassical correlation.

It should be noted that a class of maps useful for the purpose is not
limited to the one working inside $M_d(\mathbbm{C})$ apart from EnCE maps.
Furthermore, one may seek for some linear operation which is not
described by a map and has a natural extension in the dimension.
Therefore, the result shown in this paper should not be regarded as a
limitation in the use of general linear operations although it is,
of course, a natural choice to explore nonlinear EnCE maps.

Given a general nonlinear (NL) map $\Lambda_{\rm NL}$, there is a
freedom to define $I\otimes \Lambda_{\rm NL}$ under the present
circumstance where no axiom is commonly known. It is natural to impose
the condition that
$(I\otimes \Lambda_{\rm NL})(A\otimes B)=A\otimes\Lambda_{\rm NL}B$
for any matrices $A$ and $B$ in the spaces on which $I$ and
$\Lambda_{\rm NL}$ are respectively acting.
In addition, as for the case where $\Lambda_{\rm NL}$ is an nnHP nonlinear
EnCE map ${\tilde{\Theta}}_{\rm NL}$, it is desirable to impose the condition
that the map $I\otimes {\tilde{\Theta}}_{\rm NL}$ should preserve the set of
eigenvalues of any density matrix with a product eigenbasis in a similar manner
as linear one since we aim to use the map to detect and quantify nonclassical
correlation.
These conditions are satisfied
by the particular nonlinear EnCE map defined in Ref.\ \cite{SRN11}. It
is hoped that further discussions will be made and a natural axiom for
extending the dimension by a tensor product will be found so as to
utilize other nonlinear preservers for the study of nonclassical
correlation.

\nonumsection{Acknowledgement}
\noindent
A.S. and M.N. are supported by the ``Open Research Center'' Project
for Private Universities: matching fund subsidy from MEXT.
A.S. is supported by the Grant-in-Aid for Scientific
Research from JSPS (Grant No. 21800077).
R.R. is supported by Industry Canada and CIFAR.

\nonumsection{References}
\noindent
\bibliographystyle{unsrt-mod}
\bibliography{refs_nonclassical}

\begin{thebibliography}{10}

\bibitem{B99}
C.~H. Bennett, D.~P. DiVincenzo, C.~A. Fuchs, T.~Mor, E.~Rains, P.~W. Shor,
  J.~A. Smolin, and W.~K. Wootters~(1999), {\it Quantum nonlocality without
  entanglement}, Phys.\ Rev.\ A, 59, pp. 1070--1091.

\bibitem{Z02}
H.~Ollivier and W.~H. Zurek~(2001), {\it Quantum discord: A measure of the
  quantumness of correlations}, Phys.\ Rev.\ Lett., 88, pp. 017901--1--4.

\bibitem{O02}
J.~Oppenheim, M.~Horodecki, P.~Horodecki, and R.~Horodecki~(2002), {\it
  Thermodynamical approach to quantifying quantum correlations}, Phys.\ Rev.\
  Lett., 89, pp. 180402--1--4.

\bibitem{BW92}
C.~H. Bennett and S.~J. Wiesner~(1992), {\it Communication via one- and
  two-particle operators on {Einstein-Podolsky-Rosen} states}, Phys. Rev.
  Lett., 69, pp. 2881–288.

\bibitem{R06}
R.~Rahimi, K.~Takeda, M.~Ozawa, and M.~Kitagawa~(2006), {\it Entanglement
  witness derived from nmr superdense coding}, J. Phys. A: Math. Gen., 39, pp.
  2151--2159.

\bibitem{jS04}
J.~Shimamura~(2004), {\em Playing games in quantum realm}, PhD thesis, Osaka
  University, Toyonaka.

\bibitem{OjS04}
\c{S}. K.~\"{O}zdemir, J.~Shimamura, and N.~Imoto~(2004), {\it Quantum
  advantage does not survive in the presence of a corrupt source: optimal
  strategies in simultaneous move games}, Phys. Lett. A, 325, pp. 104--111.

\bibitem{SRNgame09}
A.~SaiToh, R.~Rahimi, and M.~Nakahara (2009), {\it Yet another framework for
  quantum simultaneous noncooperative bimatrix games}, In M. Nakahara, Y. Ota,
  R. Rahimi, Y. Kondo, and M. Tada-Umezaki, editors, {\em Molecular
  Realizations of Quantum Computing 2007}, World Scientific (Singapore), pp.
  223-242.

\bibitem{NT10}
A.~Nawaz and A.~H. Toor, {\it Quantum games and quantum discord}, {\rm arXiv:
  1012.1428 (quant-ph)}.

\bibitem{KL98}
E.~Knill and R.~Laflamme~(1998), {\it Power of one bit of quantum information},
  Phys.\ Rev.\ Lett., 81, pp. 5672--5675.

\bibitem{DSC08}
A.~Datta, A.~Shaji, and C.~M. Caves~(2008), {\it Quantum discord and the power
  of one qubit}, Phys.\ Rev.\ Lett., 100, pp. 050502--1--4.

\bibitem{H05}
M.~Horodecki, P.~Horodecki, R.~Horodecki, J.~Oppenheim, A.~Sen(De), U.~Sen, and
  B.~Synak-Radtke~(2005), {\it Local versus nonlocal information in
  quantum-information theory: Formalism and phenomena}, Phys.\ Rev.\ A, 71, pp.
  062307--1--25.

\bibitem{PHH08}
M.~Piani, P.~Horodecki, and R.~Horodecki~(2008), {\it No-local-broadcasting
  theorem for multipartite quantum correlations}, Phys.\ Rev.\ Lett., 100, pp.
  090502--1--4.

\bibitem{SRN08}
A.~SaiToh, R.~Rahimi, and M.~Nakahara~(2008), {\it Nonclassical correlation in
  a multipartite quantum system: Two measures and evaluation}, Phys.\ Rev.\ A,
  77, pp. 052101--1--9.

\bibitem{LF10}
S.~Luo and S.~Fu~(2010), {\it Geometric measure of quantum discord}, Phys. Rev.
  A, 82, pp. 034302--1--4.

\bibitem{G07}
B.~Groisman, D.~Kenigsberg, and T.~Mor, {\it "{Q}uantumness" versus
  "classicality" of quantum states}, {\rm quant-ph/0703103}.

\bibitem{L08}
S.~Luo~(2008), {\it Using measurement-induced disturbance to characterize
  correlations as classical or quantum}, Phys.\ Rev.\ A, 77, pp. 022301--1--5.

\bibitem{SRN11}
A.~SaiToh, R.~Rahimi, and M.~Nakahara~(2011), {\it Mathematical framework for
  detection and quantification of nonclassical correlation}, Quantum\ Inf.\
  Comput., 11, pp. 0167--0180.

\bibitem{SRN10-2}
A.~SaiToh, R.~Rahimi, and M.~Nakahara~(2011), {\it Tractable measure of
  nonclassical correlation using density matrix truncations}, Quantum Inf.
  Process., 10, pp. 431--447.

\bibitem{DVB10}
B.~Daki\'{c}, V.~Vedral, and \v{C}. Brukner~(2010), {\it Necessary and
  sufficient condition for nonzero quantum discord}, Phys. Rev. Lett., 105, pp.
  190502--1--4.

\bibitem{CCMV11}
L.~Chen, E.~Chitambar, K.~Modi, and G.~Vacanti~(2011), {\it Detecting
  multipartite classical states and their resemblances}, Phys. Rev. A, 83, pp.
  020101(R)--1--4.

\bibitem{P96}
A.~Peres~(1996), {\it Separability criterion for density matrices}, Phys.\
  Rev.\ Lett., 77, pp. 1413--1415.

\bibitem{H96}
M.~Horodecki, P.~Horodecki, and R.~Horodecki~(1996), {\it Separability of mixed
  states: necessary and sufficient conditions}, Phys.\ Lett.\ A, 223, pp. 1--8.

\bibitem{H97}
P.~Horodecki~(1997), {\it Separability criterion and inseparable mixed states
  with positive partial transposition}, Phys.\ Lett.\ A, 232, pp. 333--339.

\bibitem{SRN08-2}
A.~SaiToh, R.~Rahimi, and M.~Nakahara~(2008), {\it Evaluating measures of
  nonclassical correlation in a multipartite quantum system}, Int. J. Quant.
  Inf., 6 {\rm (Supp. 1)}, pp. 787--793.

\bibitem{W59}
E.~P. Wigner~(1959), {\em Group Theory}, Academic Press~(New York).

\bibitem{Wig}
E.~P. Wigner~(1960), {\it Phenomenological distinction between unitary and
  antiunitary symmetry operators}, J.\ Math.\ Phys., 1, pp. 414--416.

\bibitem{Wig2}
E.~P. Wigner~(1960), {\it Normal form of antiunitary operators}, J.\ Math.\
  Phys., 1, pp. 409--413.

\bibitem{SA90}
C.~S. Sharma and D.~F. Almeida~(1990), {\it A direct proof of wigner's theorem
  on maps which preserve transition probabilities between pure states of
  quantum systems}, Ann. Phys. (N.Y.), 197, pp. 300--309.

\bibitem{Mar71}
M.~Marcus~(1971), {\it Linear transformations on matrices}, J. Res. Nat. Bureau
  Stand. B. Math. Sci., 75B, pp. 107--113.

\bibitem{LT92}
C.-K. Li and N.-K. Tsing~(1992), {\it Linear preserver problems: A brief
  introduction and some special techniques}, Linear Algebra Appl., 162-164, pp.
  217--235.

\bibitem{LP01}
C.-K. Li and S.~Pierce~(2001), {\it Linear preserver problems}, Amer. Math.
  Monthly, 108, pp. 591--605.

\bibitem{MM59}
M.~Marcus and B.~N. Moyls~(1959), {\it Linear transformations on algebras of
  matrices}, Canad. J. Math., 11, pp. 61--66.

\bibitem{Minc74}
H.~Minc~(1974), {\it Linear transformations on nonnegative matrices}, Linear
  Algebra Appl., 9, pp. 149--153.

\bibitem{Minc88}
H.~Minc~(1988), {\em Nonnegative Matrices}, Wiley~(New York).

\bibitem{LT85}
P.~Lancaster and M.~Tismenetsky~(1985), {\em The Theory of Matrices (2nd Ed.)},
  Academic Press~(San Diego).

\end{thebibliography}

\end{document}